\documentstyle{article}
\font\tenrm=cmr10
\font\tenit=cmti10
\font\elevenbf=cmbx10 scaled\magstep 1
\font\elevenrm=cmr10 scaled\magstep 1

\renewenvironment{thebibliography}[1]
 { \elevenrm
   \begin{list}{\arabic{enumi}.}
    {\usecounter{enumi}     \setlength{\parsep}{0pt}
     \setlength{\itemsep}{3pt} \settowidth{\labelwidth}{#1.}
     \sloppy
    }}{\end{list}}

\begin{document}
\begin{center}{\elevenbf
THE BUBBLES OF MATTER FROM MULTISKYRMIONS.}\\
\vglue 0.5cm
{\tenrm V.B.Kopeliovich \\}
{\tenit Institute for Nuclear Research of the Russian Academy of
Sciences, Moscow 117312, Russia\\}
\end{center}
\vglue 0.3cm
{\rightskip=2pc
 \leftskip=2pc
\tenrm\baselineskip=11pt
\noindent
Approximate analytical solutions describing the skyrmions given by rational 
map ansaetze are obtained. At large baryon numbers these solutions are similar 
to the domain wall, or
to spherical bubbles with energy and baryon number density concentrated at 
their boundary. Rigorous upper bound is obtained for the masses of $RM$ 
multiskyrmions which is remarkably close to known masses, especially at
large $B$. The main properties of bubbles of matter are obtained for arbitrary
number of flavors.
\vglue 0.6cm}
\elevenrm
\baselineskip=14pt
{\bf 1.} 
Among many known soliton models used in different fields of physics
the chiral soliton approach, starting with few basic 
concepts and ingredients incorporated in the model lagrangian \cite{1,2}
provides, probably, the most realistic description of baryons and baryonic
systems. The latter appear within this approach 
as quantized solitonic solutions of equations of motion, characterized by the
so called winding number or topological charge which is identified with
baryon number $B$.
Numerical studies have shown that the chiral field configurations of lowest
energy possess different topological 
properties - the shape of the mass and $B$-number distribution - for different 
values of $B$. It is a sphere for $B=1$ hedgehog \cite{1}, a torus for $B=2$,
tetrahedron for $B=3$, cube for $B=4$, and higher
polyhedrons for greater baryon numbers. The symmetries of
various configurations for $B$ up to $22$ and their masses have been 
determined in \cite{3} (the references to earlier original papers can be found 
in \cite{3,4}). These configurations have one-shell structure and for 
$B > 6$ all of them, except two cases, are
formed from $12$ pentagons and $2B-14$ hexagons, in carbon chemistry
similar structures are known as fullerenes \cite{3}c.

The so called rational map $(RM)$ ansatz, proposed for $SU(2)$ skyrmions in 
\cite{5} and
widely used now allows to simplify the problem of finding the configurations
of lowest energy. For the $RM$ ansatz the minimization of the skyrmions energy
functional proceeds in two steps: at first step the map from $S^2 \to S^2$ is
minimized for $SU(2)$ model (for $SU(N)$ model it is a map from $S^2 \to 
CP^{N-1}$, \cite{6}), and, second, the energy functional depending on skyrmion 
profile as a function of distance from center of skyrmion is minimized.
As will be shown here, just the second step can be done analytically with
quite good accuracy. Many important properties of $RM$ multiskyrmions can
be studied in this way, and some of them do not
depend on result of the first step. This allows to make certain conclusions 
for arbitrary large $B$ and for any number of flavors $N_F=N$ independently on 
presence of numerical calculations.

So far, the chiral soliton models have been considered as a special class of 
models. Their connection with other soliton models would be of
interest, and this is also an issue of present paper.\\

{\bf 2.}
Here we consider the $SU(N)$ multiskyrmions given by rational map ansaetze; 
detailed comparison of analytical results
with numerical calculations is made in $SU(2)$ model and also in $SU(3)$ 
variant using the projector ansatz \cite{6}.
In $SU(2)$ model the chiral fields are functions of profile $f$ and unit vector
$\vec{n}$, according to definition of the unitary matrix $U \in SU(2)$
$U=c_f+is_f\vec{n}\vec{\tau}$.
For the ansatz based on rational maps the profile
$f$ depends only on variable $r$, and components of vector $\vec{n}$ - on
angular variables $\theta, \, \phi$. 
$n_1=(2\, Re\, R)/(1+|R|^2) ,\; n_2=(2\, Im\, R)/(1+|R|^2) ,\; n_3 =(1-|R|^2)/
(1+|R|^2)$, where $ R$ is a rational function of variable $z=tg(\theta/2)
exp(i\phi)$ defining the map of degree ${\cal N}$ from $S^2 \to S^2$.

The notations are used \cite{5}
$$ {\cal N} = {1\over 8\pi}\int r^2(\partial_i \vec{n})^2 d\Omega =
{1\over 4\pi}\int \frac{2i dR d\bar{R}}{(1+|R|^2)^2} $$
$${\cal I} = {1 \over 4\pi}\int r^4\frac{[\vec{\partial}n_1\vec{\partial}n_2]^2}
{n_3^2} d\Omega = {1 \over 4\pi} \int \Biggl(\frac{(1+|z|^2)}{(1+|R|^2)}
{|dR| \over |dz|}\Biggr)^4 \frac{2i dz d\bar{z}}{(1+|z|^2)^2}, \eqno (1) $$
where $\Omega$ is a spherical angle.
For $B=1$ hedgehog ${\cal N}={\cal I} =1$. ${\cal N}=B$ for configurations of
lowest energy.

The classical mass of skyrmion for $RM$ ansatz in universal 
units $3\pi^2 F_\pi/e$ is \cite{5,6}:
$$ M={1 \over 3\pi}\int \biggl\{A_N r^2f'^2 +2Bs_f^2(f'^2+1) + 
{\cal I} \frac{s_f^4}{r^2} \biggr\} dr,  \eqno (2) $$
$r$ measured in units $2/(F_\pi e)$,
where we inserted the coefficient $A_N = 2(N-1)/N$ for symmetry group $SU(N)$ 
\cite{6} to have a possibility to consider models with arbitrary number of
flavors $N=N_F$ - essentially nonembeddings of $SU(2)$ in $SU(N)$.
The expressions for the quantities ${\cal N}, {\cal I}$ for projector ansatz in
$SU(N)$ model are presented in \cite{6}.
To find the minimal energy configuration at fixed
${\cal N}=B$ one minimizes ${\cal I}$, and then finds the profile $f(r)$ by 
minimizing energy $(2)$. 
The inequality takes place: ${\cal I} \geq B^2$ \cite{5,6}.
Direct numerical calculations have shown, and the analytical treatment
here supports, that at large $B$ and, hence, large ${\cal I}$ multiskyrmion looks
like a spherical bubble with profile equal to $f=\pi$ inside and $f=0$
outside. The energy and $B$-number density of this configuration is concentrated 
at its boundary, similar to the domain walls system considered in \cite{7}
in connection with cosmological problems.

Denote $\phi=cos f$, then the energy $(2)$ can be presented as
$$M={1\over 3\pi}\int \biggl\{{1 \over (1-\phi^2)}
\biggl[A_Nr^2\phi'^2+2B(1-\phi^2)^2\biggr] +2B\phi'^2+{\cal I}(1-\phi^2)^2/r^2
\biggr\}dr,  \eqno (3) $$
with $\phi$ changing from $-1$ at $r=0$ to $1$ at $r\to\infty$. 
The first half of $(3)$ is the second order term contribution into the mass,
the second - the Skyrme term contribution. At fixed $r=r_0$ the latter 
coincides exactly with $1$-dimensional domain wall energy. 
It is possible to write the second order contribution in $(2)$ in the form:
$$M^{(2)}={1\over 3\pi}\int \biggl\{{A_Nr^2 \over (1-\phi^2)}
\biggl[\phi'- \sqrt{{2B\over A_N}}(1-\phi^2)/r\biggr]^2 + 2r\sqrt{2A_NB}\phi' 
\biggr\}dr,  \eqno (4) $$
and similar for the $4$-th order Skyrme term. The equality $\phi'=\sqrt{2B/A_N}
(1-\phi^2)/r $ eliminates considerable part of integrand in $(4)$.
Therefore, it is natural to consider function $\phi$ satisfying the following
differential equation:
$$ \phi'= {b \over 2r}(1-\phi^2) \eqno (5) $$
which has solution satisfying boundary conditions $\phi(0)=-1$ and 
$\phi(\infty )=1$:
$$ \phi = \frac{(r/r_0)^b-1}{(r/r_0)^b+1} \eqno (6)$$
with arbitrary $r_0$ - the distance from the origin of the point where 
$\phi=0$ and profile $f=\pi/2$. $r_0$ can be considered as a radius of
multiskyrmion.

After substitution of this ansatz one obtains the soliton mass
in the form:
$$M(B,b) = {1 \over 3\pi}\int \biggl\{(A_Nb^2/4+2B)(1-\phi^2)
+ (Bb^2/2+{\cal I})(1-\phi^2)^2/r^2\biggr\} dr \eqno (7) $$
Integrating over $dr$ using known
expressions for the Euler-type integrals, e.g. 
$$\int_0^\infty \frac{dr}{1+(r/r_0)^b}=\frac{r_0\pi}{b\, sin(\pi/b)},
\qquad b>1 $$ 
allows to obtain the mass of multiskyrmion in simple analytical form as a 
function of parameters $b$ and $r_0$:
$$M(B,r_0,b)= \frac{4}{3b^2sin(\pi/b)}\biggl[(A_Nb^2/4+2B)r_0 +{1\over 3r_0}
(Bb^2+2{\cal I})(1-1/b^2) \biggr] \eqno (8)$$
which gives after simple minimization in $r_0$
$$r_0^{min}=2 \biggl[\frac{(Bb^2+2{\cal I})(1-1/b^2)}{3(A_Nb^2+8B)}
\biggr]^{1/2} \eqno (9) $$
and
$$M(B,b)/B= \frac{4}{3b\,sin(\pi/b)}\bigl[(b^2+2{\cal I}/B)(A_N b^2+8B)
(1-1/b^2)/(3b^2B)\bigr]^{1/2} \eqno (10) $$
For any value of the power $b$  $(10)$ provides an upper bound for the mass
of $RM$ skyrmion. To get better bound we should minimize $(10)$ in $b$.
At large enough $B$ when it is possible to neglect the influence of slowly
varying factors $(1-1/b^2)$ and $b\,sin(\pi/b)$ we obtain easily that
$$b_{min}=b_0= 2({\cal I}/A_N)^{1/4},\qquad
r_0^{min}\simeq \biggl[{2\over 3}\biggl(\sqrt{\frac{{\cal I}}{A_N}}-
{1\over 4}\biggr)\biggr]^{1/2} \eqno (11)$$
and, therefore,
$${1\over 3}(2+\sqrt{{\cal I}A_N}/B) < {M\over B} \leq {1 \over 3}(2 + 
\sqrt{{\cal I}A_N}/B)
\frac{4}{b_0 sin(\pi/b_0)}\biggl[{2\over 3}\biggl(1-
{1\over 4}\sqrt{A_N/{\cal I}}\biggr)\biggr]^{1/2}. \eqno (12) $$
The lower bound in $(12)$ was known previously \cite{5,6}.
The correction to the value $b_0$ can be found including into minimization
procedure the factor $(1-1/b^2)$ and variation of $b\, sin (\pi/b)\simeq 
\pi [1-\pi^2/(6b^2)] $.
It provides:
$$\delta \, b=\frac{B(\pi^2/3-1)(2+\sqrt{A_N{\cal I}}/B)^2}
{16 {\cal I}^{3/4}A_N^{1/4}}, 
\eqno (13) $$
and the value $b=b_0+\delta\,b $ should be inserted into $(10)$. This improves
the values of $M/B$ for $B=1,2,3... $ but provides negligible effect for 
$b \sim 17$ and greater, since $\delta\,b \sim 1/\sqrt{B}$.
The comparison of numerical calculation result and analytical upper bound
$(12)$ is presented in the {\bf Table}.
\newpage
\begin{center}
\begin{tabular}{|l|l|l|l|l|l|l|l|l|l|l|l|}
\hline
$B$ &$2$ &$3 $& $4$ &$5 $ &$6 $&$7$&$13$&$17$&$22$&$32$&$64$ \\
\hline
$M/B|_{RM}$  &$1.208$&$1.184$&$1.137$&$1.147$&$1.137$&$1.107$&
$1.098$&$1.092$&$1.092$&---&--- \\
\hline
$b(B)$&$3.89$&$4.47$&$4.85$&$5.39$&$5.80$&$6.03$&$8.00$&$9.02$&$10.23$&$12.24$&
$17.2$\\
\hline
$M/B|_{appr}$&$1.229$&$1.198$&$1.151$&$1.158$&$1.147$ &$1.117$&$1.106$&
$1.0976$&$1.098$&$1.094$&$1.089$ \\
\hline
$M/B|_{num}$&$1.1791$&$1.1462$&$1.1201$&$1.1172$&$1.1079$&$1.0947$&
$1.0834$&$1.0774$&$1.0766$&---&--- \\
\hline
\hline
$M/B|^{SU_3}_{RM}$ &$1.222$&$1.215$&$1.184$&$1.164$&$1.146$&---&---&---&---&---
&--- \\
\hline
$b(B)^{SU_3}$&$3.57$&$4.08$&$4.47$&$4.83$&$5.13$&$5.46$&$7.13$&$8.06$&$9.09$&
$10.86$&$15.19$ \\
\hline
$M/B|^{SU_3}_{appr}$&$1.259$&$1.231$&$1.198$&$1.176$&$1.156$&$1.149$&$1.127$&
$1.121$&$1.116$&$1.111$&$1.106$ \\
\hline
\end{tabular}
\end{center}
{\tenrm\baselineskip=11pt The skyrmion mass per unit $B$-number in universal 
units $3\pi^2F_\pi/e$
for $RM$ configurations, approximate and precise solutions. The approximate 
values are calculated using
formula $(10)$ with the power $b=b_0+\delta\,b$. The numerical values for 
$SU(2)$ model are from
the papers \cite{3} and earlier papers. The last $3$ lines show the result for $SU(3)$ projector 
ansatz \cite{6} and approximation to this case, $A_N=4/3$ .}\\

Numerically $(10)-(13)$ provide upper bound which differs from the
masses of all known $RM$ skyrmions within $\sim 2$\%, beginning with $B=4$,
see Table. 
Even for $B=1$, where the method evidently should not work well, we obtain
$M=1.289$ instead of precise value $M=1.232$.
For maximal values of $B$ between $17$ and $22$ where the value of ${\cal I}$ 
is calculated, the upper bound exceeds the $RM$ value of mass by $0.5 \%$ only.
We took here the ratio $R_{I/B}={\cal I}/B^2$ in the cases where
this ratio is not determined yet, the same as for highest $B$ where it is known,
i.e. $1.28$ for $SU(2)$ case\cite{3}, $B=32$ and $64$, and $1.037$ for $B>6$ in $SU(3)$
\cite{6}. For $R_{I/B}=1$ the numbers in {\bf Table} decrease by $\sim 0.1\%$
only in the latter case.
Note also, that asymptotically at large $B$ the ratio of upper and lower bounds
$$ {M_{max}\over M_{min}} = {4\over \pi} \biggl({2\over 3}\biggr)^{1/2} 
\simeq 1.0396,\eqno (14) $$
i.e. the gap between upper and lower bounds is less than $4 \%$,
independently on $B$, the particular value of ${\cal I}$ and the number of 
flavors $N$.
With decreasing ${\cal I}$ the upper bound decreases proportionally to
the lower bound. In view of such good quantitative agreement of analytical and
numerical results the studies of basic properties of bubbles of matter made in 
present paper are quite reliable.

The width (or thickness) $W$ of the bubble shell can be estimated easily. We 
can define the half-width as a distance between points where $\phi = \pm 1/2$, then:
$$ W = 4\biggl[{2 \over 3}\biggl(1- {1 \over b_0^2}\biggr)\biggr]^{1/2} ln 3.
\eqno (15) $$
It is clear that at large $B \; W$ is constant, and does not depend on the
number of flavors $N$ as well. The radius of the bubble
$(11)$ grows with increasing $B$ like $[{\cal I}/A_N]^{1/4} $.

Previously we considered large $B$ skyrmion within the "inclined step" 
approximation \cite{4}.
Let $W$ be the width of the step, and $r_0$ - the radius of the 
skyrmion where the profile $f =\pi/2$. $f= \pi/2 - (r-r_o)\pi/W$
for $r_o-W/2 \leq r \leq r_o+W/2 $.
This approximation describes the usual domain wall energy \cite{7}
with accuracy $\sim 9.5$\% .

We wrote the energy in terms of $W, r_0$, then minimized it
with respect to both of these parameters, and find the minimal value of
energy.
$$ M(W, r_0)={\pi^2 \over W}(B+A_Nr_0^2)+W \biggl(B+
\frac{3 {\cal I}}{8r_0^2}\biggr) \eqno (16) $$
This gave
$$W_{min}=\pi\biggl[\frac{B+A_Nr_0^2}{B+3{\cal I}/(8r_0^2)}
\biggr]^{1/2} \eqno (17) $$
and, after minimization, $r^2_{0\,min}=\sqrt{3 {\cal I} /(8 A_N)}\simeq
0.612 \sqrt{{\cal I}/A_N}$, close to the above result $r_{0\,min}^2 \simeq
0.667 \sqrt{{\cal I}/A_N}$.
In dimensional units $r_0= (6{\cal I}/A_N)^{1/4} / (F_\pi e)$.
Since $ {\cal I} \geq B^2 $, the radius of minimized configuration grows
as $\sqrt{B}$, at least.
$W_{min}=\pi$, i.e. it does not depend on $B$ for any $SU(N)$, similar to
previous result $(15)$ which gives $W\simeq 3.59$ for large B, all in units
$2/(F_\pi e)$.
The energy
$$M_{min}\simeq (2B+\sqrt{3A_N {\cal I} /2} ) /3 \eqno (18) $$
In difference from previous result $(12)$, $(18)$ does not give the upper
bound for the skyrmion mass, and for small $B$, indeed, the value of $(18)$ is
smaller than calculated masses of skyrmions.
For $SU(2)$ model $A_N=1$ and the energy $M_{min}= (2B+\sqrt{3{\cal I} /2})/3$.
The formula gives the numbers for $B=3,..., \, 22$ in agreement
with calculation within $RM$ approximation within $2-3 \%$ \cite{3,5}.

More detailed analytical calculation made here confirms the results of such 
"toy model" approximation and both reproduce the picture of $RM$ skyrmions as 
a two-phase object, a spherical bubble with profile $f=\pi$ inside and $f=0$ outside,
and a fixed thickness shell with fixed surface energy density,
$\rho_M^{surf} \simeq (2B+\sqrt{3{\cal I}A_N /2 })/ (12\pi r_0^2) $.
The average volume density of mass in the shell is
$$\rho_M^{vol}={3\pi \over 64W}(2B+\sqrt{{\cal I}A_N})\sqrt{A_N/{\cal I}}
F_\pi^4 e^2 \eqno (19) $$
and for $SU(2)$ model at large $B$ it is approximately equal to $\sim 0.3
Gev/Fm^3$ at reasonable choice of model parameters $F_\pi=0.186 \,Gev,\; 
e=4.12$ \cite{4}, i.e. about twice greater than normal density of nuclei.

{\bf 3.} Consider also the influence of the chiral symmetry breaking mass term 
which is described by the lagrangian density
$$ M.t. =\tilde{m} \int r^2(1-cos\,f) dr , \qquad cos\,f=\phi \eqno (20) $$
$\tilde{m} = 8\mu^2/(3\pi F_\pi^2 e^2) $, $\mu=m_\pi$. For strangeness, charm, or bottom
the masses $m_K$, $m_D$ or $m_B$ can be inserted instead of $m_\pi$.

Instead of above expression $(8)$ we obtain now
$$ M(B,r_0,b) = \alpha (B,b) r_0 +\beta (B,b)/r_0+ m \,r_o^3  \eqno (21)$$
with $\alpha,\,\beta$ given in $(8)$ and $m=2\pi\tilde{m}/(b\,sin(3\pi/b))$.
It is possible to obtain precise minimal value of the mass
$$M(B,b)={2r_0^{min} \over 3}\bigl(\sqrt{\alpha^2+12m\beta}+2\alpha\bigr)
\eqno(22) $$
at the value of $r_0$
$$r_0^{min}(B,b) =\biggl[\frac{\sqrt{\alpha^2 +12\,m\beta}-\alpha}{6m}
\biggr]^{1/2} \eqno (23) $$
When the mass $m$ is small enough, the expansion in $12m\beta/\alpha^2$ can be
made, and one obtains the following reduction of dimension $r_0$:
$$ r_0 \to r_0 - {3m\over 2\alpha} \biggl(\frac{\beta}{\alpha}\biggr)^{3/2}
\simeq \sqrt{{2\over 3}}\biggl(\frac{{\cal I}}{A_N}\biggr)^{1/4}\biggl[1-\,
\frac{3\pi m}{2(2B+\sqrt{{\cal I}A_N})}\biggl(\frac{{\cal I}}{A_N}\biggr)
^{3/4}\biggr], \eqno (24) $$
and increase of the soliton mass
$$\delta M = M \frac{m\beta}{2\alpha^2}\biggl[1-\frac{9m\beta}{8\alpha^2}
\biggr]\simeq M \,m\frac{\pi ({\cal I}A_N)^{3/4}}{2(2B+\sqrt{{\cal I}A_N}).}
 \eqno (25) $$
We used that at large $B$
$$\alpha \simeq {1\over 3\pi}\biggl(A_Nb+{8B\over b}\biggr), \qquad
\beta \simeq {4\over 9\pi}\biggl(Bb + \frac{2{\cal I}}{b}\biggr). \eqno (26) $$
As it was expected from general grounds, dimensions of the soliton decrease
with increasing $m$.
However, even for large value of $m$ the structure of multiskyrmion at 
large $B$
remains the same: the chiral symmetry broken phase inside of the spherical
wall where the main contribution to the mass and topological charge is 
concentrated. 
The value of the mass density inside of
the bubble is defined completely by the mass term with $1-\phi= 2$.
The baryon number density distribution is quite similar, with only difference
that inside the bag it equals to zero.
It follows from these results that $RM$ approximated multiskyrmions cannot
model real nuclei at large $B$, probably $B > 12 -20$, and configurations like
skyrmion crystals \cite{8} may be more valid for this purpose.

It is of interest to study what happens at very large values of the mass,
when $12 m\beta \gg \alpha^2 $.
Making expansion in $\alpha^2/12m\beta$ we obtain:
$$r_0^{min} \simeq [\beta/(3m)]^{1/4}, \qquad
 M_{m \to \infty}(B,b) \simeq {4\over 3} (3m\beta^3)^{1/4}. 
\eqno (27) $$
Minimizing $M(B,b)$ in $b$ gives $b_0^{min} = \sqrt{2{\cal I}/B}, \;
\beta^{min}= 8\sqrt{2{\cal I}B}/(9\pi) $ and
$$ M_{m\to \infty}(B) \simeq {16 \over 9}[4\sqrt{2}m/(3\pi^3)]^{1/4}
[B{\cal I}]^{3/8}. \eqno (28) $$
So, in this extreme case $b \sim \sqrt{B}$, $r_0\sim (\mu)^{-1/2} B^{3/8}$,
the mass of the soliton increases as
$M\sim \sqrt{\mu} B^{9/8}$, 
and the binding should become weaker with increasing baryon number of skyrmion.

It is possible also to calculate analytically the tensors of inertia of 
multiskyrmion configurations within this approximation, see \cite{4} for
definitions and formulas. \\

{\bf 4.}
By means of consideration of a class of functions $(6)$ we established the 
link between toplogical soliton models in rational maps approximation and
the soliton models of "domain wall" or "spherical bubble" type.
Of course, it is a domain wall of special kind, similar to honeycomb.
The upper bound for the energy of multiskyrmions is obtained which is very
close to the known energies of $RM$ multiskyrmions, especially at largest $B$,
and is higher than the known lower bound by $\sim 4\%$ only.

The following properties of bubbles of matter from $RM$ multiskyrmions are
established analytically, mostly independent on particular values of the
quantity ${\cal I}$:

The dimensions of the bubble grow with $B$ as $\sqrt{B}$, or as 
${\cal I}^{1/4}$, whereas the mass is proportional to $\sim B$ at large $B$.
Dimensions of the bubble decrease slightly with increasing $N$ - the
number of flavors, $r_0 \sim [N/\bigl(2(N-1)\bigr)]^{1/4}$,
see $(11),(12)$.

The thickness of the bubbles envelope $(15)$ is constant at large $B$ and does 
not depend on the number of flavors, therefore, the average
surface mass density is constant at large $B$, as well as average volume 
density of the shell. Both densities increase slightly with increasing $N$.
At the same time the mass and $B$-number
densities of the whole bubble $ \to 0$ when $B\to \infty$, and this is in 
contradiction with nuclear physics data. 

It follows from the above consideration that the spherical bubble or bag 
configuration can be obtained from
the lagrangian written in terms of chiral degrees of freedom only, i.e.
the Skyrme model lagrangian leads at large baryon numbers to the formation 
of spherical bubbles of matter, i.e. it provides a field-theoretical
realization of the bag-type model. 
This picture of mass distribution in $RM$ multiskyrmions condradicts at first 
sight to what is known about nuclei,
however, it emphazises the role of periphery of the nucleus and could be an
argument in favour of shell-type models of nuclei. The skyrmion crystals 
\cite{8} are believed to be more adequite for modelling nuclear matter.

It would be of interest to perform the investigation of the dynamics of
bubbles in chiral soliton models similar to that performed recently for
the simplified two-component sigma model, or sine-Gordon model in $(3+1)$ 
dimensions \cite{9}.
Observations concerning the structure of large $B$ multiskyrmions made here
can be useful in view of possible cosmological applications of Skyrme-type
models, see e.g. \cite{10}. The large scale structure of
the mass distribution in the Universe \cite{11} is similar to that in 
topological soliton models, and it can be the consequence of the similarity of 
the laws in micro- and macroworld.

I am indebted to P.M. Sutcliffe for informing me about the results of
\cite{3}b,c before publication and to W.J.Zakrzewski for interest in
the questions discussed in present paper.
The work is supported by RFBR grant 01-02-16615. \\

{\elevenbf\noindent References}
\vglue 0.1cm

\end{document}